\documentclass[]{spie}  %

\usepackage{amsmath,amsfonts,amssymb}
\usepackage{graphicx}
\usepackage[table,dvipsnames]{xcolor}

\usepackage[colorlinks=true, allcolors=blue]{hyperref}

\newcommand{\submm}{\mbox{(sub-)mm}}
\newcommand{\Submm}{\mbox{(Sub-)mm}}

\title{The key science drivers for the Atacama Large Aperture Submillimeter Telescope (AtLAST)}

\author[a]{Mark Booth}
\author[a]{Pamela Klaassen}
\author[b]{Claudia Cicone}
\author[c]{Tony Mroczkowski}
\author[b, d]{Sven Wedemeyer}
\author[e, f, g]{Kazunori Akiyama}
\author[h]{Geoffrey Bower}
\author[i]{Martin A. Cordiner}
\author[j, k, l, m]{Luca Di Mascolo}
\author[n, o]{Doug Johnstone}
\author[c]{Eelco van Kampen}
\author[p, q]{Minju M. Lee}
\author[r, s]{Daizhong Liu}
\author[t]{John Orlowski-Scherer}
\author[u, v]{Amélie Saintonge}
\author[w]{Matthew Smith}
\author[x]{Alexander E. Thelen}

\affil[a]{UK Astronomy Technology Centre, Royal Observatory Edinburgh, Blackford Hill, Edinburgh EH9 3HJ, UK}
\affil[b]{Institute of Theoretical Astrophysics, University of Oslo, PO Box 1029, Blindern 0315, Oslo, Norway}
\affil[c]{European Southern Observatory (ESO), Karl-Schwarzschild-Strasse 2, Garching 85748, Germany}
\affil[d]{Rosseland Centre for Solar Physics, Institute of Theoretical Astrophysics, University of Oslo, Postboks 1029 Blindern, N-0315 Oslo, Norway}
\affil[e]{Haystack Observatory, Massachusetts Institute of Technology, 99 Millstone Road, Westford, MA 01886, USA}
\affil[f]{National Astronomical Observatory of Japan, 2-21-1 Osawa, Mitaka, Tokyo 181-8588, Japan}
\affil[g]{Black Hole Initiative, Center for Astrophysics $\mid$ Harvard \& Smithsonian, 20 Garden Street, Cambridge, MA 02138, USA}
\affil[h]{Academia Sinica Institute of Astronomy and Astrophysics, 645 N. A'ohoku Place, Hilo, HI 96720, USA}
\affil[i]{Astrochemistry Laboratory, Code 691, NASA Goddard Space Flight Center, Greenbelt, MD 20771, USA.}
\affil[j]{Laboratoire Lagrange, Université Côte d'Azur, Observatoire de la Côte d'Azur, CNRS, Blvd de l'Observatoire, CS 34229, 06304 Nice cedex 4, France}
\affil[k]{Astronomy Unit, Department of Physics, University of Trieste, via Tiepolo 11, Trieste 34131, Italy}
\affil[l]{INAF -- Osservatorio Astronomico di Trieste, via Tiepolo 11, Trieste 34131, Italy}
\affil[m]{IFPU -- Institute for Fundamental Physics of the Universe, Via Beirut 2, 34014 Trieste, Italy}
\affil[n]{NRC Herzberg Astronomy and Astrophysics, 5071 West Saanich Rd, Victoria, BC, V9E 2E7, Canada}
\affil[o]{Department of Physics and Astronomy, University of Victoria, Victoria, BC, V8P 5C2, Canada}
\affil[p]{Cosmic Dawn Center (DAWN), Denmark}
\affil[q]{DTU-Space, Technical University of Denmark, Elektrovej 327, DK2800 Kgs. Lyngby, Denmark}
\affil[r]{Max-Planck-Institut f\"{u}r extraterrestrische Physik, Giessenbachstrasse 1 Garching, Bayern, D-85748, Germany}
\affil[s]{Purple Mountain Observatory, Chinese Academy of Sciences, 10 Yuanhua Road, Nanjing 210008, China}
\affil[t]{Department of Physics and Astronomy, University of Pennsylvania, 209 South 33rd Street, Philadelphia, PA, 19104, USA}
\affil[u]{Department of Physics and Astronomy, University College London, Gower Street, London WC1E 6BT, UK}
\affil[v]{Max-Planck-Institut f\"ur Radioastronomie (MPIfR), Auf dem H\"ugel 69, D-53121 Bonn, Germany}
\affil[w]{School of Physics \& Astronomy, Cardiff University, The Parade, Cardiff CF24 3AA, UK}
\affil[x]{Division of Geological and Planetary Sciences, California Institute of Technology, Pasadena, CA 91125, USA.}

\authorinfo{Further author information: \\M. Booth: E-mail: mark.booth@stfc.ac.uk}
\begin{document} 
\maketitle

\begin{abstract} %
Sub-mm and mm wavelengths provide a unique view of the Universe, from the gas and dust that fills and surrounds galaxies to the chromosphere of our own Sun. Current single-dish facilities have presented a tantalising view of the brightest (sub-)mm sources, and interferometers have provided the exquisite resolution necessary to analyse the details in small fields, but there are still many open questions that cannot be answered with current facilities: \textit{Where are all the baryons? How do structures interact with their environments? What does the time-varying (sub-)mm sky look like?} In order to make major advances on these questions and others, what is needed now is a facility capable of rapidly mapping the sky spatially, spectrally, and temporally, which can only be done by a high throughput, single-dish observatory. An extensive design study for this new facility is currently being undertaken. In this paper, we focus on the key science drivers and the requirements they place on the observatory. As a 50m single dish telescope with a 1--2$^\circ$ field of view, the strength of the Atacama Large Aperture Submillimeter Telescope (AtLAST) is in science where a large field of view, highly multiplexed instrumentation and sensitivity to faint large-scale structure is important. AtLAST aims to be a sustainable, upgradeable, multipurpose facility that will deliver orders of magnitude increases in sensitivity and mapping speeds over current and planned telescopes. 
\end{abstract}


\section{INTRODUCTION}
\label{intro}  %
Sub-mm and mm (hereafter \submm) wavelengths provide us with a view of a wide variety of astrophysical sources, from the gas and dust that permeates and surrounds galaxies and the hot gas in galaxy clusters (via the Sunyaev-Zeldovich (SZ) effect) to the atmospheres of the Sun and planets in our own Solar system.

Single-dish telescopes have been the foundation of \submm{} astronomy, conducting large extra-galactic surveys\cite{Reuter2020}, discovering new types of galaxies\cite{Blain2002}, mapping our own Galaxy\cite{Schuller2009} and providing some of the first resolved images of debris discs\cite{Holland1998}. From this foundation,  interferometers, such as the Submillimeter Array, Northern Extended Millimeter Array and Atacama Large Millimeter/submillimeter Array (ALMA), emerged providing drastic increases in resolution, for probing detailed structure of a wide variety of astrophysical targets. But these improvements in resolution come at the cost of missing flux in extended structures \cite{Bonanomi2024}. Interferometers are also in danger of becoming source-starved without large scale surveys to help direct them at where to point next. With these limitations in mind, what is needed now is a telescope capable of reaching sensitivities comparable to those of ALMA, at resolutions of a few arcseconds whilst retaining the sensitivity to large-scale structure and with a degree-scale field of view (FoV).

In pursuit of this goal, there is currently a design study underway for a 50m class single-dish telescope operating at \submm{} wavelengths in the form of the  Atacama Large Aperture Submillimeter Telescope (AtLAST)\footnote{\url{https://atlast.uio.no}} project\cite{Klaassen2020}. This design study addresses the governance, telescope design, site selection, operations, sustainability and science drivers of the observatory. The current reference design of the telescope is described in Ref. \citenum{Mroczkowski2024}. The telescope is being designed with sustainability in mind, including an innovative regenerative breaking system to reduce power consumption\cite{Kiselev2024}. The options for powering the observatory with renewable energy are being investigated in detail, both from the technological\cite{Viole2023} and sociological\cite{Valenzuela-Venegas2023} points of view, including considerations on the full life-cycle assessment of the energy system \cite{Viole2024,Viole2024b}.

\begin{table}
    \centering
    \begin{tabular}{|p{1.6cm}|p{4.5cm}|p{4.5cm}|p{4.5cm}|}
    \hline
\rowcolor{Goldenrod}
\multicolumn{4}{|c|}{\bf Key Science Drivers for AtLAST}\\
    \hline
     {\cellcolor{Goldenrod} ~}
        &  \textit{\textbf{Where are all the baryons?}} &  
      \textit{\textbf{How do structures interact with their environments?}}& 
      \textit{\textbf{What does the time-varying \submm{} sky look like?}}\\ 
    \hline
         {\cellcolor{Goldenrod} \bf Detailed science goal} &  Measuring the total gas and dust content of the Milky Way and other galaxies, in the interstellar, circumgalactic, and intergalactic media, reaching down to the sensitivities required to probe the typical populations of sub-mm sources.&  Understanding the lifecycle of gas and dust near and far; mapping the baryon cycle on multiple-scales; observing the interplay between gravity, radiation, turbulence, magnetic fields, and chemistry and their mutual feedback.  
& Identifying the mechanisms responsible for time variability across astrophysical sources: from the Solar corona and other objects in our solar system to luminosity bursts in everything from protostars to active galactic nuclei.\\
\hline
{\cellcolor{Goldenrod} \bf Detailed technical specification}&  
    \textbf{\textit{High sensitivity to the faint signals}} (at sub-mK levels) on \textit{\textbf{large scales}} ($\geq$1 deg$^2$) from even the most diffuse and cold gas through sub-mm line tracers. Wide field ($>$500 deg$^2$)  continuum surveys capturing the plane of our galaxy and resolving 82\% of the cosmic infrared background, probing typical populations and looking back over 95\% of the age of the Universe. &  
    \textit{\textbf{High spectral resolution}} and \textit{\textbf{polarisation }}measurements on the relevant size scales for cores (0.1 pc, our galaxy), clumps (10 pc, nearby galaxies) and cloud complexes ($\sim$ few kpc, distant universe) to quantify the chemistry, disentangle the dynamics, and measure the magnetic fields working together to shape the evolution of structures within their larger-scale environments. &
    An operations model that allows for \textit{\textbf{highly cadenced and rapid response observations}} and data reduction pipelines with in-built \textit{\textbf{transient detection algorithms}}\textbf{; }high time-resolution (few seconds) observations of our Sun and other stars.\\
    \hline
    \end{tabular}
    \caption{The three key science themes identified where AtLAST can have a transformative impact.}
    \label{tab:key_questions}
\end{table}

In this paper we focus on the key science drivers for the observatory as derived from consultation with the scientific community (in Section \ref{science}) and the requirements they place on the observatory (in Section \ref{require}). A first step in this process was the organisation of an initial consultation with the community that solicited use cases for the telescope\cite{Ramasawmy2022}. Following this, a number of science working groups (SWGs) were set-up to investigate the use cases in further detail to provide specific, quantified estimations of what will be possible with AtLAST. A summary of the results of this exercise are presented here, with the key science drivers derived for the observatory presented in Table \ref{tab:key_questions}.

\section{SCIENCE DRIVERS}
\label{science}

The science achievable with a large aperture \submm{} single dish telescope spans a wide range of astrophysical fields. Through the SWGs, we produced a series of white papers on the topics of surveying the high redshift Universe\cite{vanKampen2024}, the SZ effect\cite{DiMascolo2024}, the circumgalactic medium (CGM) of galaxies \cite{Lee2024}, nearby galaxies\cite{Liu2024}, our Galaxy\cite{Klaassen2024}, the Solar System\cite{Cordiner2024}, the Sun\cite{Wedemeyer2024},  and the more general topic of time-domain observations\cite{Orlowski-Scherer2024}.  %
Here we summarise the key science presented in these white papers, and what they mean for telescope requirements along with details on the topic of \submm{} science with very long baseline interferometry (VLBI)\cite{Akiyama2023}. For the white papers, all time estimates are derived using the AtLAST Sensitivity Calculator\footnote{\url{https://senscalc.atlast.uio.no} or \url{https://github.com/ukatc/AtLAST_sensitivity_calculator}\cite{Klaassen2024b}}.

\subsection{Distant Universe}
\textit{See van Kampen et al.\ (2024)\cite{vanKampen2024} for further details.}

Observations at \submm{} wavelengths provide a view of the dust and gas in galaxies across cosmic time, including the cold molecular phase that can be uniquely probed in this band. Dust obscures the light from stars within a galaxy, hindering observations in the ultraviolet and optical, yet by reprocessing such radiation, the emission from dust and gas can inform us of star formation processes throughout the evolution of galaxies. With current \submm{} single-dish facilities we are limited to observing the brightest of these galaxies.
In order to study the rest of the population 
and accurately determine both their number counts and redshift distribution, a large aperture telescope is needed to drastically lower the confusion limit\footnote{The confusion limit characterises the ability to distinguish one source from another within the resolution element and so effectively translates into a sensitivity threshold.} along with a large FoV to rapidly map large areas of the sky. With, for example, 1000 hours of observing time on AtLAST, covering a 1000~deg$^2$ at 350~$\mu$m, we expect to reach a 3$\sigma$ limit of 570~$\mu$Jy. With this limit, 82\% of the Cosmic Infrared Background will be resolved into individual sources, allowing the detection of over 50 million galaxies in continuum. Taking a 1~deg$^2$ portion of this (containing around 100,000 galaxies), we can follow up with a spectroscopic survey to map the evolution of the cold gas content across cosmic time. In addition, line-intensity mapping (tomography) with AtLAST allows stringent limits on various cosmological parameters, including the Hubble constant. In addition to this, a distinct large survey of distant galaxy clusters with AtLAST will produce a high-redshift counterpart to local large surveys of rich clusters like the well-studied Abell catalogue, to understand the impact of environment on the formation and evolution of these distant cluster galaxies. 

\subsection{The Warm and Hot Universe}
\textit{See Di Mascolo et al.\ (2024)\cite{DiMascolo2024} for further details.}

   \begin{figure} %
   \begin{center}
   \begin{tabular}{c} 
   \includegraphics[width=0.95\textwidth]{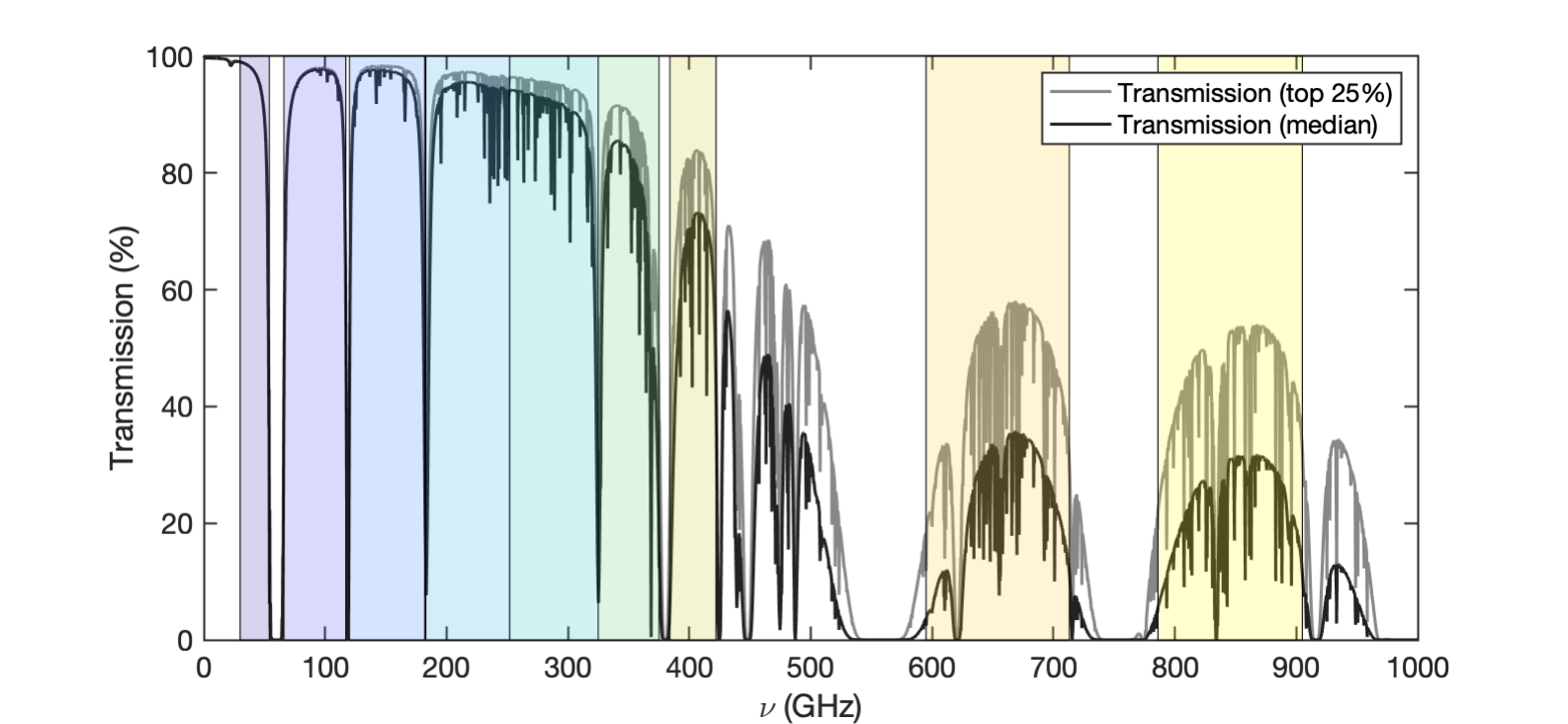}
	\end{tabular}
	\end{center}
   \caption[example] 
   { \label{fig:bands} 
Atmospheric transmission at the Chajnantor Plateau. Eight possible continuum bands, recommended in Ref. \citenum{DiMascolo2024}, that would span the frequency range of AtLAST are shown by the shaded regions. Figure adapted from Ref. \citenum{DiMascolo2024}.}
   \end{figure} 

Hot gaseous atmospheres, such as those permeating galaxy clusters, groups, and individual galaxies, can be studied via the SZ effect,\cite{Sunyaev1970, Sunyaev1972, Sunyaev1980} which is a spectral distortion of the cosmic microwave background (CMB) through inverse Compton scattering by high energy electrons in said atmosphere. 
The SZ effect provides a redshift independent view on the thermal and kinetic properties of the warm/hot ($>10^5 \, \rm K$) ionised gas, which is an important component of the intracluster medium (ICM) but is also expected to fill the circumgalactic medium (CGM) of galaxies undergoing powerful feedback. Studying this gas provides essential information for constraining the co-evolution of galactic populations and large-scale cosmological structures, for tracing the matter assembly in the Universe and its thermal history. Given the large scales of interest for SZ studies, a sensitive single dish with a degree-scale FoV is necessary to avoid filtering out the SZ signal and to provide the mapping speed necessary to rapidly cover large areas of the sky. Multiple wavebands spanning, for example, 30--905~GHz (see Fig.~\ref{fig:bands}) are needed to disentangle the various SZ components and avoid foreground and background contamination from the cosmic infrared background and dust in the cluster itself. The high resolution and sensitivity of AtLAST could enable the detection of the SZ effect in high-z protoclusters\cite{DiMascolo2023} and in individual massive galaxies. 

\subsection{Circumgalactic Medium (CGM)} 
\textit{See Lee et al.\ (2024)\cite{Lee2024} for further details.}

Galaxies are found to be surrounded by a reservoir of gas and dust extending beyond the interstellar medium (ISM), commonly referred to as the circumgalactic medium (CGM). Understanding the CGM allows us to evaluate the feedback and feeding mechanisms that impact the galaxy's ability to sustain star formation. These feedback mechanisms leave observable signatures in the density, temperature, metallicity, and morphology of the CGM.  The total mass of the CGM reservoir and the contribution from different gas phases (neutral/cold versus highly ionised/hot) that exist within it are largely unconstrained, as are the physical processes that affect and shape the CGM. Observing the CGM is challenging because of its primarily faint, diffuse, and extended nature. To observe it requires a telescope with sensitivity to these types of large-scale structures (the emission can extend to many hundreds of kiloparsecs -- corresponding to an angular size of degrees for galaxies within 10~Mpc of the Milky Way), with the ability to resolve sub-kiloparsec scale clumps and a dynamic range capable of distinguishing the CGM from the significantly brighter ISM. Simulations\cite{Schimek2024} demonstrate that the [CII] 158$\mu$m, [OIII] 88$\mu$m, [CI] 609$\mu$m, 370 $\mu$m, and CO spectral lines are ideal for studying the CGM gas. Observing these lines requires a broad wavelength range (from far-infrared to mm) to cover galaxies at all redshifts, as well as an extremely high sensitivity to large-scale emission to capture their faint signal from the high-z CGM. To achieve these goals requires stable spectral baselines to properly analyse the broad line wings produced by powerful galactic outflows, which one one of the main mechanisms transporting gas and energy from ISM to CGM scales\cite{Cicone18,Cicone21}.

\subsection{Nearby galaxies}
\textit{See Liu et al.\ (2024)\cite{Liu2024} for further details.}

Nearby galaxies present an opportunity to investigate the ISM, star formation, chemical, magnetic and dynamical physics across a galaxy and down to scales within individual giant molecular clouds (GMCs). The diversity amongst them allows us to better understand galactic physics in environments different to the Milky Way. For instance, the Large and Small Magellanic Clouds (LMC \& SMC) are the nearest laboratories for studying star formation at low metallicity where cooling differs from that in our galaxy -- important for extrapolating to high-$z$ galaxies with similar or even lower metalicities. What is cooling the gas and dust? How are structures able to collapse? How do they feedback into their larger environments?  A challenge with observing the Magellanic clouds is that they are so nearby and extended (several degrees in the sky) that interferometers cannot fully recover the extended gas and dust structures, and current generation single dish telescopes cannot reach the sensitivities required to map the faint gas and dust emission. To image the full dust content of the Magellanic clouds down to a sensitivity of 1.5~$\mu$Jy/beam, 22~$\mu$Jy/beam and 19~$\mu$Jy/beam at 100~GHz, 400~GHz and 680~GHz respectively, across the full field would require of order 1000 hours. 
With a multi-band highly multiplexed continuum camera, significant gains can be made into our understanding of how dust properties vary across different environments by studying these systems and others in the local group.The addition of polarimetric observations of these galaxies means that not only can we trace the faint diffuse emission, but we can link it to the magnetic fields permeating these regions and quantify the balance between magnetic, gravitational, turbulent and other forces working in these galaxies. Mapping the full (few deg$^2$) extent of the dust in the magellanic clouds, for instance, to a depth where measuring 1\% polarisation fractions at each position would take of order 3000 hours, and is impossible with current or planned facilities. Magnetic fields play an important role in star formation,  regulating its efficiency, as well as the more general dynamics of the ISM, but there is still a lack of understanding in how they originate, are amplified or are affected by star formation activity in general in external galaxies.  Using spectroscopy, the molecular gas in the local Universe can be measured in great detail, enabling the determination of gas masses, surface densities, gas densities, temperatures and chemistry in a far greater number of galaxies than previously studied. 

\subsection{Our Galaxy}
\textit{See Klaassen et al.\ (2024)\cite{Klaassen2024} for further details.}

The Milky Way provides us with the most detailed view of galactic scale processes from the magnetic field structure that permeates the ISM to the formation and evolution of stars and their planetary systems.  Studies of the relationship between gravity, turbulence, magnetic fields and feedback must be considered as a whole if we are to understand star formation within its broader ecosystem (e.g. cores, clumps, fillaments, GMCs). With polarised continuum and spectroscopy we can map the magnetic fields, dynamics, temperature and chemistry of the plane of the Milky Way down to a resolution capable of resolving individual star forming cores (0.1 pc) up to 10~kpc away. High spectral resolution ($\leq$0.1~km\,s$^{-1}$) is necessary to spectrally resolve the dynamics of the gas as traced by ionised and molecular spectral lines. From these cores, planetary systems develop. \Submm{} observations of the circumstellar discs of these systems provide information on the evolution of planetary systems and can even act as indicators of where planets currently are in a system. Young stellar objects (YSOs) are still in their natal environments and at distances where large area surveys are necessary to study the evolution from protoplanetary to debris discs of the full star forming region (e.g. Taurus subtends 30$^\circ$ on the sky) and thus necessitate a large FoV to efficiently survey them. Debris discs around nearby stars, on the other hand, are so close that they can reach arcminutes in size, meaning that to discover discs as faint as our own Kuiper belt requires a highly sensitive single-dish telescope.

\subsection{Solar System}
\textit{See Cordiner et al.\ (2024)\cite{Cordiner2024} for further details.}

Within our own planetary system, planets, moons and individual comets are all detectable in the \submm. Spectroscopy of the giant planets teaches us about their chemistry (and therefore, their origins) and the dynamics of their atmospheres. Spectroscopy of these bodies teaches us about their habitability and the availability of the chemical ingredients necessary for life. Current single-dish observatories have limitations in terms of sensitivity, dynamic range and spatial coverage needed for further advances in planetary science; interferometers lack sensitivity to large scales and the constantly changing configuration can be problematic for time critical observations and tracking both short-term and seasonal changes. With a 50~m dish, Venus, Mars, Jupiter and Saturn can all be resolved throughout their orbits and Uranus and Neptune can also be resolved at the shortest wavelengths (see for instance Figure 1 of Ref \citenum{Cordiner2024}). In order to study their chemistry, a spectroscopic dynamic range of $\sim10^5$ is needed due to the continuum brightness distribution of the sources. The moon Titan also presents a tantalising target given its terrestrial-like atmosphere, where deep observations have the potential to detect new isotopologues of known molecules that may be the key to understanding the long-term physico-chemical evolution of the atmosphere. Icy moons, such as Enceladus, are known to produce plumes of gases emanating from the ocean underneath ice. Studying these with \submm{} spectroscopy provides an avenue for assessing the chemistry of the ocean. Comets are thought to be largely unchanged since their formation in the young Solar System and so studying their chemical composition provides insights into the chemistry of the early Solar System. %

\subsection{The Sun}
\textit{See Wedemeyer et al.\ (2024)\cite{Wedemeyer2024} for further details.}

\Submm\ wavelengths provide insights into the Solar atmosphere, particularly its chromosphere. A deeper understanding of the chromosphere teaches us about the three dimensional structure of the Sun, heating of the corona, origin of solar winds, drivers of solar activity and space weather. The Sun is highly dynamic and so particularly calls for repeat observations with a variety of temporal cadences. Short observations with a cadence of a few seconds are necessary to study solar flares, repeat observations over days can show the evolution of sunspots and daily scans of the whole Sun allow us to observe the evolution over a solar rotation (if repeated for a month) or even over a solar activity cycle (if continued for 11 years). Multi-wavelength observations are highly valuable to probe the different layers of the atmosphere and these need to be taken simultaneously if we want to be able to study the temporal evolution of the layers e.g. during a flare, and AtLAST could bridge the gap between optical and radio images of that evolution. The importance of the magnetic field of the Sun requires that any continuum camera designed to observe the Sun should have polarimetric capabilities.

\subsection{Transients and variability}
\textit{See Orlowski-Scherer et al.\ (2024)\cite{Orlowski-Scherer2024} for further details.}

There are a wide variety of transient and variable phenomena detectable at \submm{} wavelengths, including (proto)stellar flares, novae, active galactic nuclei and massive black hole binaries.
Study of transients and variability in the \submm{} is still in its infancy. This is largely due to the need for a combination of both high sensitivity, so that more than just the brightest events can be detected, and a wide field of view, to increase the possibility of time variable phenomena being captured within the field. AtLAST will provide both of these and thus greatly increases the possibilities for dedicated (such as stellar and protostellar flares) and commensal (such as asteroids passing through or background gamma ray bursts) observations. 
In order to make the most of such observations, special consideration must be paid towards development of specialised data products, reduction pipelines and the policies for sharing such data. To facilitate variability studies with repeated observations, a high standard of flux calibration ($\sim$1-3\%) must be maintained. Additionally, for transient events seen with AtLAST there will be the need to trigger observations at other telescopes and similarly, for AtLAST to perform follow-up of transient events seen by other telescopes.

\subsection{Very long baseline interferometry}
\textit{See Akiyama et al.\ (2023)\cite{Akiyama2023} for further details.}

The images of M87$^*$ and Sgr~A$^*$ produced by the Event Horizon Telescope (EHT) demonstrated the power of very long baseline interferometry (VLBI) at mm wavelengths. Pushing this into the sub-mm will produce revolutionary benefits in terms of resolution. Observing at 690~GHz will provide the resolution necessary to resolve the individual photon rings produced by different physical mechanisms that make up the emission seen in the prior EHT observations while also enabling the observations of a much greater set of targets. There are currently very few telescopes capable of observing at these wavelengths and so if AtLAST is included in the EHT, it will add significantly to the capabilities of the EHT, particularly given its high sensitivity comparable to that of ALMA: being a single-dish, it has significant benefits as an EHT station over an interferometer due to the ability to implement multi-frequency observations that allow order of magnitude improvements in sensitivity via frequency phase transfer.

\section{TECHNICAL REQUIREMENTS}
\label{require}
 \begin{table}[]
    \centering
    \begin{tabular}{l|l|p{9cm}}
        Parameter & Goal & Key science drivers \\
        \hline
        Primary mirror diameter & $\geq$50~m & All cases \\
        Field of view & $\geq$1~deg$^2$ & SZ effect, Milky Way's Galactic plane, surveys of high-redshift galaxies, CGM around nearby galaxies \\
        Scan speed & $\sim$1~deg\,s$^{-1}$ & The Sun, transients \\
        Angular resolution at 1~mm & $<5''$ & All cases \\
        Wavelength range & 0.3-3~mm & All cases \\[2mm]
        Continuum polarisation & \checkmark & Magnetic fields in galaxies and the Sun \\
        Continuum bandwidth & 64 GHz+ & Most cases\\
        Continuum multi-chroic & $>3$ bands & Transients, dust in our Galaxy, and the Sun \\
        Continuum pixels & 10$^5$-10$^6$ & Most cases\\
        [2mm]
        Spectral resolution & 0.01~km\,s$^{-1}$ & Dynamics and chemistry of our Galaxy \\
        Spectral bandwidth & 64~GHz & Planetary and cometary atmospheres \\
        Spectral pixels & $100-1000$ & Our Galaxy, Nearby galaxies\\
        Stable spectral baselines & \checkmark & CGM \\[2mm]
        Solar observing & \checkmark & The Sun \\
    \end{tabular}
    \caption{Summary of the technical requirements and key science drivers for each.}
    \label{trequire}
\end{table}

The science drivers presented here place a variety of stringent requirements on the telescope, its instrumentation, operations and data reduction facilities. A summary of these requirements and the key drivers for each can be found in Table \ref{trequire}.

\subsection{Telescope Requirements}

The key science goals described in this paper require the resolution and collecting area delivered by a 50-m dish with the high sensitivity to large scale structures that comes from a filled and large FoV.
As noted above, these goals cannot be accomplished by current \submm{} facilities but specifically require the next generation capabilitites provided by AtLAST.

A FoV of at least 1~deg$^2$ is needed in order not to filter out (via scanning) the degree-scale structures such as those relevant to studying the warm/hot gas in galaxy clusters via the SZ effect, the cold atomic and molecular emission from the CGM in nearby galaxies as well as the plane of the Milky Way (covering $>500$ deg$^2$), and to efficiently perform  wide-field extragalactic surveys that are needed to gather statistics over several 100s of deg$^2$.

\begin{figure} %
   \begin{center}
   \begin{tabular}{c} 
   \includegraphics[width=0.95\textwidth]{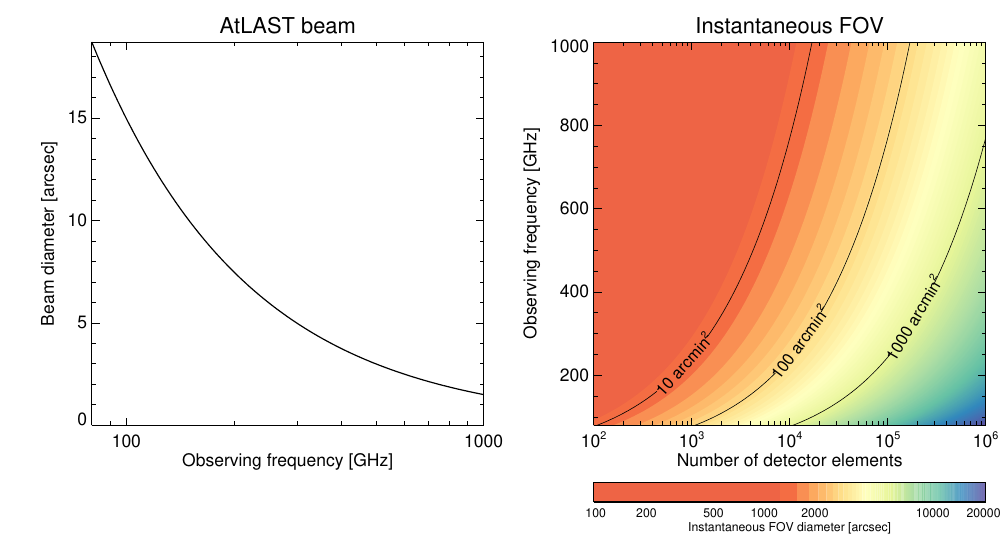}
	\end{tabular}
	\end{center}
   \caption[] 
   { \label{fig:beam} 
\emph{Left:} Beam diameter for a 50m telescope as a function of observing frequency. \emph{Right:} Instantaneous FoV as a function of the observing frequency and number of detector pixels, assuming a circular packing fraction of 80\%. The resulting area of the FoV is shown by the black contours. Figure adapted from Ref. \citenum{Wedemeyer2024}.}
   \end{figure}

In addition to a large FoV, these projects also push for an angular resolution on the order of arcseconds at \submm{} wavelengths. This makes it possible to resolve molecular clouds around nearby galaxies, star-forming cores in our Galaxy, and the terrestrial and giant planets of our own Solar System. This resolution also ensures that confusion between targets can be minimised. In particular, this reduces the extragalactic confusion limit to below the sensitivity required to detect normal star forming galaxies at z$\sim$10\cite{Blain2002,vanKampen2024}. This also enables observations of moons in the Solar System by ensuring that they can easily be resolved from the planets they orbit which, without high-dynamic-range in the instrumentation, can saturate detectors before the moon is detected.
All of this is achievable with a 50~m dish that results in resolutions of $<5''$ at all sub-mm wavelengths (frequencies $>300$~GHz; see the left plot of Figure \ref{fig:beam}). A dish of this size also makes it possible to reach high sensitivities in a short time, e.g. reaching $\sim20\,\mu$Jy/beam at 850~$\mu$m in one hour\cite{DiMascolo2024} (c.f. $\sim1.7$~mJy/beam for the Submillimetre Common-User Bolometer Array 2 instrument on the James Clerk Maxwell Telescope\cite{Holland2013}). This will enable the detection of galaxies as faint as our own Milky Way at high redshift, and planetesimal belts as faint as our own Kuiper belt around the nearest stars.

Cold dust has a peak wavelength in the far-infrared to sub-mm depending on redshift; therefore the sub-mm windows provide the best way to observe this dust from the ground. The sub-mm windows also provide the possibility of observing a host of spectral lines from atomic, ionised and molecular gas species which tell us about the dynamics, ionisation state and chemistry of the gas. Observing at sub-mm wavelengths results in a better resolution than in the mm, providing all the benefits noted above. 
To observe at these high frequencies, a high and dry site is essential along with a high surface accuracy.

Target of opportunity follow-up of events triggered by other observatories can require rapid follow-up e.g. to find the source of a gravitational wave event or the afterglow of a gamma ray burst. Scan speeds on the order of a degree per second are necessary to fully capture and characterise such events.

\subsection{Instrumentation Requirements}
The science goals listed here require a mix of continuum and spectral line observations including polarisation preserving and solar observing capabilities. In addition, a dedicated multi-frequency instrument will be required to enable VLBI science.

Large format continuum cameras will make the most of the large FoV with expected pixel counts of order 10$^5-10^6$ (see the right plot of Figure \ref{fig:beam}) in the first generation. A wide wavelength coverage from 350~$\mu$m down to at least 3~mm is desirable to get a broad sampling of the SEDs of the targets. For transient phenomena, such as solar flares and GRBs, being able to simultaneously observe multiple (at least 3) wavelengths is necessary to understand the temporal evolution of the SED, while for other cases, more pixels in a single band is preferable for covering larger areas per observation. A broad bandwidth, covering the full atmospheric window in each band is needed to maximise the sensitivity (see Figure \ref{fig:bands}). %

For observing spectral lines, both high-resolution spectrometers and large format integral field units (IFUs) are envisaged. For understanding the chemistry of our Galaxy, a spectral resolution of 0.01~km\,s$^{-1}$ is necessary to ensure that spectral line shapes can be resolved in the coldest cores and a spectrometer bandwidth of at least 16~GHz enables the simultaneous observation of multiple emission lines needed for chemical surveys. As with the continuum camera, a large number (ideally 100s-1000s) of pixels are required to reduce mapping times. Planetary and cometary atmospheres contain a wide variety of molecules, thus pushing for as much as 64~GHz bandwidth in order to capture as many as possible. A high spectroscopic dynamic range of $\gtrsim10^5$ will be needed for planetary atmospheres to distinguish weak atmospheric spectral lines from the bright continuum. For surveys of local galaxies, a large FoV is more important than spectral resolution and so an IFU with a 1~deg$^2$ FoV and at least 32~GHz bandwidth will also be necessary. Stable spectral baselines are required for the broad lines expected in external galaxies due to outflows and CGM emission.

In order for a telescope to be part of a VLBI network, a heterodyne receiver system will be necessary and to accomplish the listed science goals this will need to have a high sensitivity, wide bandwidth and multi-frequency capabilities.

\subsection{Telescope Operations}
The wide variety of science cases presented here calls for a good mix of programmes from pointed observations of just a few hours to legacy surveys of thousands of hours. Transient and variable targets place further constraints on the operations, requiring the ability to do time critical observations, regular monitoring, targets of opportunity, coordination with other observatories and rapid follow-up. In particular, this motivates the development of an automated transient search pipeline that would enable new transients to be quickly shared with the community, allowing prompt follow-up with other observatories.

\section{SUMMARY}
\label{summary}
In this paper we present a community-driven set of science cases that all motivate the construction of a new large single-dish observatory designed to observe in the sub-mm. The science cases range from the distant Universe to the Solar System and all scales in between, motivating the requirement of a multipurpose observatory, capable of large area surveys, pointed observations and transient investigations.
Specifically, this telescope is required to be 50~m in size with a $\geq$1~deg$^2$ FoV and hosting continuum, spectral line and VLBI instrumentation with polarimetric and solar observing capabilities. These requirements are driving the design of the AtLAST observatory that is currently undergoing an initial design study. With the capability to achieve the science goals listed here and many, many more, AtLAST will stimulate the field of sub-mm astronomy for generations to come.

\acknowledgments %
This project has received funding from the European
Union’s Horizon 2020 research and innovation programme under grant agreement No.\ 951815 (AtLAST).

\bibliography{main} %
\bibliographystyle{spiebib} %

\end{document}